\documentclass[a4paper,11pt,final]{article}

\usepackage{amsmath,amssymb,bm,times,graphicx}

\usepackage[small,bf]{caption}

\usepackage{hyperref}
\hypersetup{pdftitle={Mean-field critical behaviour and ergodicity break in a nonequilibrium one-dimensional RSOS growth model}, pdfauthor={J. Ricardo G. Mendonça}, pdfsubject={Statistical mechanics, growth models, RSOS, ergodicity}, pdfkeywords={Nonequilibrium growth model, roughening transition, stationary probability density, Ginzburg-Landau, mean-field, ergodicity}, unicode, extension=pdf, pdfnewwindow, pdftoolbar, pdfmenubar, pdffitwindow=false, pdfstartview={FitH}, colorlinks, linkcolor=blue, citecolor=blue, urlcolor=blue}
\def\leq{\leqslant}

\def\I{{\rm i}}

\begin{document}

\pagestyle{empty}
\renewcommand{\thefootnote}{\fnsymbol{footnote}}

\begin{titlepage}

\begin{center}

{\Large \bf Mean-field critical behaviour and ergodicity break in \\
            a nonequilibrium one-dimensional RSOS growth model}

\vspace{6ex}

{\large {\bf J. Ricardo G. Mendon\c{c}a$^{a,b,}$}\footnote{Email: {\tt \href{mailto:jricardo@if.usp.br}{\nolinkurl{jricardo@if.usp.br}}}}}

\vspace{1ex}

{\it {$^{a}$Instituto de F\'{\i}sica, Universidade de S\~{a}o Paulo} \\ \makebox[0pt][c]{Rua do Mat\~{a}o, Travessa~R 187, Cidade Universit\'{a}ria -- 05508-090 S\~{a}o Paulo, SP, Brazil}}

{\it {$^{b}$Universidade Federal de S\~{a}o Carlos -- Campus Sorocaba} \\ \makebox[0pt][c]{Rodovia Jo\~{a}o Leme dos Santos, km 110 -- 18052-780 Sorocaba, SP, Brazil}}

\vspace{6ex}

{\large \bf Abstract \\}

\vspace{2ex}

\parbox{120mm}
{We investigate the nonequilibrium roughening transition of a one-dimensional restricted solid-on-solid model by directly sampling the stationary probability density of a suitable order parameter as the surface adsorption rate varies. The shapes of the probability density histograms suggest a typical Ginzburg-Landau scenario for the phase transition of the model, and estimates of the ``magnetic'' exponent seem to confirm its mean-field critical behaviour. We also found that the flipping times between the metastable phases of the model scale exponentially with the system size, signaling the breaking of ergodicity in the thermodynamic limit. Incidentally, we discovered that a closely related model not considered before also displays a phase transition with the same critical behaviour as the original model. Our results support the usefulness of off-critical histogram techniques in the investigation of nonequilibrium phase transitions. We also briefly discuss in an appendix a good and simple pseudo-random number generator used in our simulations.

\vspace{2ex}

{\noindent}{\bf Keywords}: Nonequilibrium growth model $\cdot$ roughening transition $\cdot$ stationary probability density $\cdot$ Ginzburg-Landau $\cdot$ mean-field exponent $\cdot$ ergodicity $\cdot$ pseudo-random number generator

\vspace{2ex}

{\noindent}{\bf PACS 2010}: 05.50.+q $\cdot$ 64.60.Cn $\cdot$ 68.35.Rh

\vspace{2ex}

{\noindent}{\bf Journal ref}.: \href{http://www.worldscinet.com/ijmpc/}{{\it Int.~J.~Mod.~Phys.~C\/} {\bf 23} (2012)}}

\end{center}

\end{titlepage}

\pagestyle{plain}


\section{\label{intro}Introduction}

During the last three decades or so, it has been found that nonequilibrium one-di\-men\-sional growth models may exhibit roughening transitions \cite{savit,wolf,aehm96,aehm98}, in contrast with the widespread lore---based on simple entropy arguments---according to which one-di\-men\-sional interfaces in thermal equilibrium are always rough \cite{landau}. This discovery helped to expose, by means of concrete examples, some fundamental differences that exist between equilibrium and nonequilium systems, e.\,g., that the dynamic fluctuations of a nonequilibrium stationary state are in general not equivalent to the thermal fluctuations of an equilibrium state. It seems that, even in the stationary state, in nonequilibrium interacting statistical systems time is more like another dimension of the system than a mere source of dynamic fluctuations, although in general it does not scale like the spatial dimensions of the system.\footnote{That time could possibly be taken as an additional dimension in a related equilibrium model seems to have been suggested first by R.~L.~Dobrushin in mid-1960s to I.~I.~Piatetski-Shapiro and coworkers, that then began numerical work on cellular automata that ultimately led to the ``positive probabilities conjecture.'' We are indebted to Professor A.~L.~Toom (UFPE) for this remark.}

In Alon {\it et al.\/} \cite{aehm96}, a one-di\-men\-sional growth model was introduced that, when the adsorption of adatoms by the surface is small, desorption of adatoms from the edges of rough droplets together with absence of desorption from smooth terraces provide local mechanisms that eliminate islands of the rough phase, thus stabilising the flat phase despite the ergodicity of the model. Local rules dissolving islands of a minority phase provide an efficient mechanism of stabilisation against noise and have had a major role in the development of the foundations of nonequilibrium statistical physics \cite{kurdyumov,toom,bennett,lebowitz,gacs}. Recently, other conditions that can possibly give rise to phase-transitions in one-di\-men\-sional systems with short-range interactions have also been investigated; we refer the reader to ref.~\cite{cuesta} for an exposition.

In this work we investigate the roughening transition of the restricted solid-on-solid (RSOS) growth model introduced in ref.~\cite{aehm96} by means of direct measurements of the stationary probability density of a suitable order parameter. In this way we were able to locate the critical point of the model and to estimate its ``magnetic'' critical exponent straighforwardly by Monte Carlo simulations. This sort of approach seems to have been introduced in ref.~\cite{arndt} and has been used since then in varied contexts \cite{pronina,popkov,clincy}. The break of ergodicity of the model in the thermodynamic limit could also be established from the flipping times between its metastable phases, which were found to scale exponentially with the system size.

This article goes as follows. In section~\ref{prob} we present the model we are interested in and our approach through the stationary probability density. In section~\ref{rough} we detail our Monte Carlo simulations and estimate the critical point and the critical exponent $\theta$ of the model. In section~\ref{desorp} we look at an extended model in which desorption from smooth terraces are allowed and answer whether this extended model displays a phase transition and spontaneous symmetry breaking. Section~\ref{flip} discusses the flipping times between the metastable phases of the model and its ergodicity. Finally, in section~\ref{summary} we summarise our conclusions and indicate some directions for further investigation.


\section{\label{prob}The RSOS growth model and the stationary probability density}

\begin{figure}
\centering
\includegraphics[viewport=103 315 389 527, scale=0.65, angle=-90, clip]{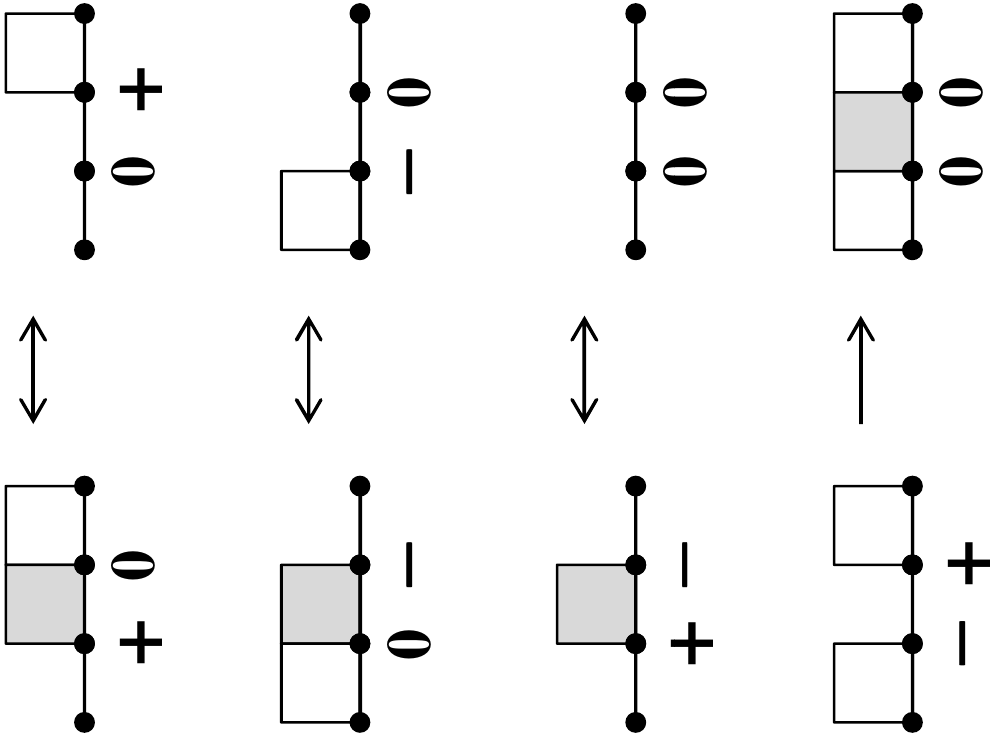}
\caption{\label{fig1}Elementary transitions in the RSOS growth model defined by the rates (\ref{gammas}). The moving adatoms are depicted in grey.}
\end{figure}

Let $h_{\ell} \in {\mathbb N}$ be the height of a surface or interface at site $\ell$ of a finite lattice $\Lambda \subset {\mathbb Z}$ with $|\Lambda|=L$ sites and periodic boundary conditions. The surface evolves by attempting, sequentially and at randomly chosen sites, adsorption of an adatom $h_{\ell} \to h_{\ell}+1$ with probability $q\,{\rm d}t$, and desorption of an adatom $h_{\ell} \to \min \{h_{\ell-1},h_{\ell}\}$ or $h_{\ell} \to \min \{h_{\ell},h_{\ell+1}\}$ each with probability $\frac{1}{2}(1-q)\,{\rm d}t$. We now impose the RSOS condition $|h_{\ell+1}-h_{\ell}| \leq 1 \  \forall\, \ell \in \Lambda$, which suggests the use of the link variables $c_{\ell} = h_{\ell+1}-h_{\ell} \in \{-1,0,1\}$. Denoting by $\Gamma^{ab}_{cd}$ the rate at which the elementary process $(a,b) \to (c,d)$ occurs, we can describe the above growth model in the links representation by the set of rates
\begin{subequations}
\label{gammas}
\begin{equation}
\label{gammas1a}
\Gamma^{+0}_{0+} = \frac{1}{2}(1-q), \qquad \Gamma^{0+}_{+0} = q,
\end{equation}
\begin{equation}
\label{gammas1b}
\Gamma^{-0}_{0-} = q, \qquad \Gamma^{0-}_{-0} = \frac{1}{2}(1-q),
\end{equation}
\begin{equation}
\label{gammas1c}
\Gamma^{+-}_{00} = 1-q, \qquad \Gamma^{00}_{+-} = q,
\end{equation}
\begin{equation}
\label{gammas1d}
\Gamma^{-+}_{00} = q.
\end{equation}
\end{subequations}
This model is translation invariant and conserves the total charge $Q^{+}-Q^{-}$, with each sector of total charge corresponding to a closed class of the stochastic process. Notice that, differently from the models investigated in refs.~\cite{arndt,clincy,kafri02,kafri03}, this model does not conserve $Q^{+}$ and $Q^{-}$ individually. The adsorption and desorption moves corresponding to the above rates are represented in figure~\ref{fig1}.

According to (\ref{gammas}), as $q$ increases creation of $+-$ as well as annihilation of $-+$ pairs increase while the remaining processes induce spatial segregation of charges. This corresponds to an increase in adsorption and in the growth of islands, leading to rougher configurations. As $q$ lowers, increased annihilation of $+-$ pairs together with more symmetric diffusion of 0's flatten the surface. It is known that a roughening transition occurs at $q_{c} \simeq 0.188$ \cite{aehm96,aehm98,jrgm}. An order parameter that captures this transition is given by
\begin{equation}
\label{order}
M_{L}=\frac{1}{L}\sum_{\ell= 1}^{L}m_{\ell}
=\frac{1}{L}\sum_{\ell= 1}^{L}(-1)^{h_{\ell}}.
\end{equation}
This non-conserved order parameter anticipates the interpretation of the roughening transition as the spontaneous break of the ${\mathbb Z}_{\infty}$ symmetry related with the invariance of the growth process under an arbitrary integer shift $h_{\ell} \to h_{\ell}+n$ in the heights, for while in the rough phase all heights are exploited evenly, in the flat phase the system spontaneously selects one level around which the heights fluctuate. We then expect $M=\lim_{L \to \infty}M_{L}$ to be finite in the flat phase while vanishing in the rough phase due to canceling fluctuations.

In this work we focus on the evaluation of the stationary probability density $P_{L}(M)$ of the order parameter $M_{L}$ defined above. Besides estimating the critical point and the main critical exponent asociated with the transition directly from $P_{L}(M)$, we also estimate the first passage time through a certain rough configuration from an initially flat configuration---the flipping time---and show that it grows exponentially with the system size, indicating that in the thermodynamic limit the process breaks ergodicity, signaling a phase transition.

A caveat is due about the approach employed in this article. Although it is tempting to define a free energy-like functional $F_{L}(M)=-\ln P_{L}(M)$ once $P_{L}(M)$ has been determined, we will avoid doing this. Firstly, because it is not necessary, since we do not intend to strugle with convexity and maximality issues here. Secondly, because it is not clear if the stationary measure can be expressed in the thermodynamic limit as a weighted exponential of some local function when detailed balance does not hold---Toom's model provides a striking counterexample \cite{toom,bennett,lebowitz}. Thus, except for one mention to a $\lambda M^{4}$ scenario in section~\ref{rough}, we avoid free energy-like arguments in what follows.


\section{\label{rough}Roughening transition}

Our Monte Carlo simulations ran as follows. The interface is initialised as a completely flat interface with all heights $h_{\ell}$ equal, i.\,e., with all $c_{\ell}=0$. In this configuration, the pseudo-spins $m_{\ell}$ should be initialised all with the same sign. It is a matter of choice to pick $+1$ or $-1$ for the initial $m_{\ell}$, because we can anchor $h_{1}(0)$ at any integer we like.

For a given value of $q$, the initial configuration is relaxed through $L^{2}$ Monte Carlo steps (MCSs), with one MCS equal to $L$ sequential attempts of movement. $M_{L}$ is then sampled every other MCS and the data accumulated as a normalised histogram $P_{L}(M)$. We drew the pseudo-random numbers used in our simulations from a combined linear congruential and 3-shift generator that is very fast, possesses a period larger than $10^{19}$, and passes several tests of randomness (cf.\ appendix~A) \cite{marsaglia}.

We obtained $P_{L}(M)$ by using relatively small lattices and sampling a relatively large number of times. In order to obtain symmetric histograms, we change the sign of the pseudo-spins $m_{\ell}$ every after we sample $M_{L}$. This allows us to sample both sides of the histogram without having to wait exponentially long times for the system to shift between them. This procedure is equivalent to simulating two systems in parallel.

\begin{figure}
\centering
\includegraphics[viewport = 68 106 523 745, scale=0.40, angle=-90, clip]{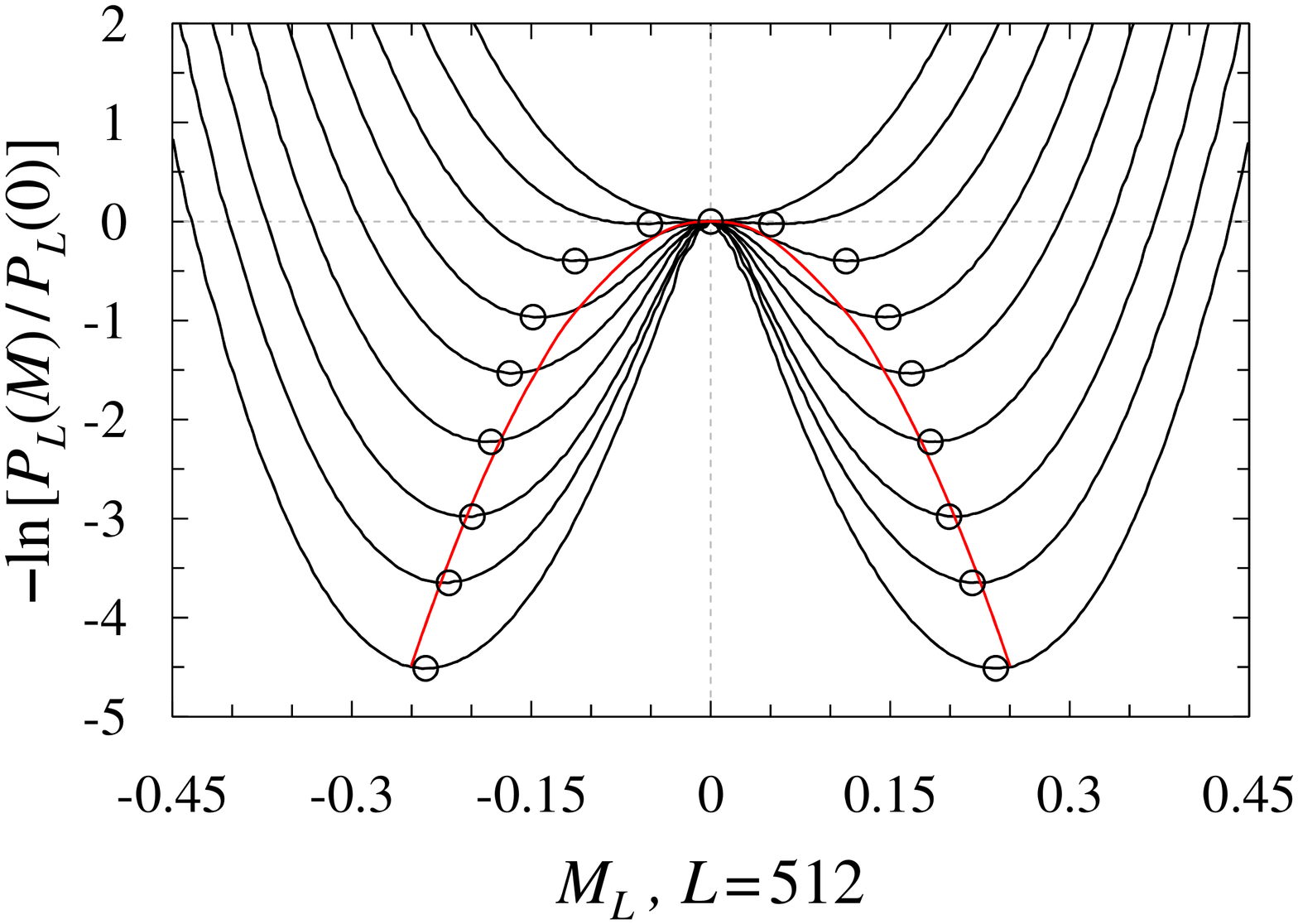}
\caption{\label{fig2}Logarithm of the stationary probability densities $P_{L}(M)$ for $L=512$ and $q$ between $0.180$ (lowermost curve) and $0.188$ (uppermost curve). Each curve was obtained from $10^{7}$ samplings of $M_{L}$. We also depict a quadratic fit ($y \simeq -71.7\,x^{2}$) to the minima of the curves (open circles), that although not being very tight, anticipates that the critical exponent $\theta$ should be close to $1/2$.}
\end{figure}

Scanning through $q$ gives the set of curves shown in figure~\ref{fig2}. This figure depicts a typical Ginzburg-Landau scenario of a second-order phase transition. Since we are dealing with a finite system, we could define $P_{L}(M) \sim \exp[-F_{L}(M)]$ with
\[
F_{L}(M) = \frac{1}{2}\mu_{L}(q)M^{2} + \frac{1}{4} \lambda_{L}(q)M^{4} + O(M^{6}),
\]
with $\mu_{L}(q) = \mu^{(0)}_{L}(q_{c}-q) + O[(q_{c}-q)^{3}]$ and $\lambda_{L}(q) > 0$. This form of $P_{L}(M)$ would predict a mean-field critical behaviour for the order parameter, $M \propto (q_{c}-q)^{\theta}$ with $\theta=\frac{1}{2}$. This prediction is supported by previous results in the literature: in ref.~\cite{aehm96}, $\theta$ was evaluated as $0.55(5)$ (the numbers between parentheses indicate the uncertainty in the last digits of the data), while for a model of yeastlike growth of fungi colonies with parallel dynamics it has been found that $\theta \simeq 0.50$, although in this case the phase transition cannot be associated with the spontaneous breaking of a symmetry \cite{lopez}. Moreover, in a certain line in the phase diagram of a closely related one-di\-men\-sional next-nearest-neighbour asymmetric exclusion process it has been found that $\theta = 0.54 \pm 0.04$ \cite{nnnasep}. In ref.~\cite{aehm98}, however, the slightly less accurate but significantly higher value $\theta = 0.66 \pm 0.06$ was published, pushing the estimate to a value that does not fit within a pure Ginzburg-Landau scenario.

\begin{figure}
\centering
\includegraphics[viewport=68 100 513 748, scale=0.40, angle=-90, clip]{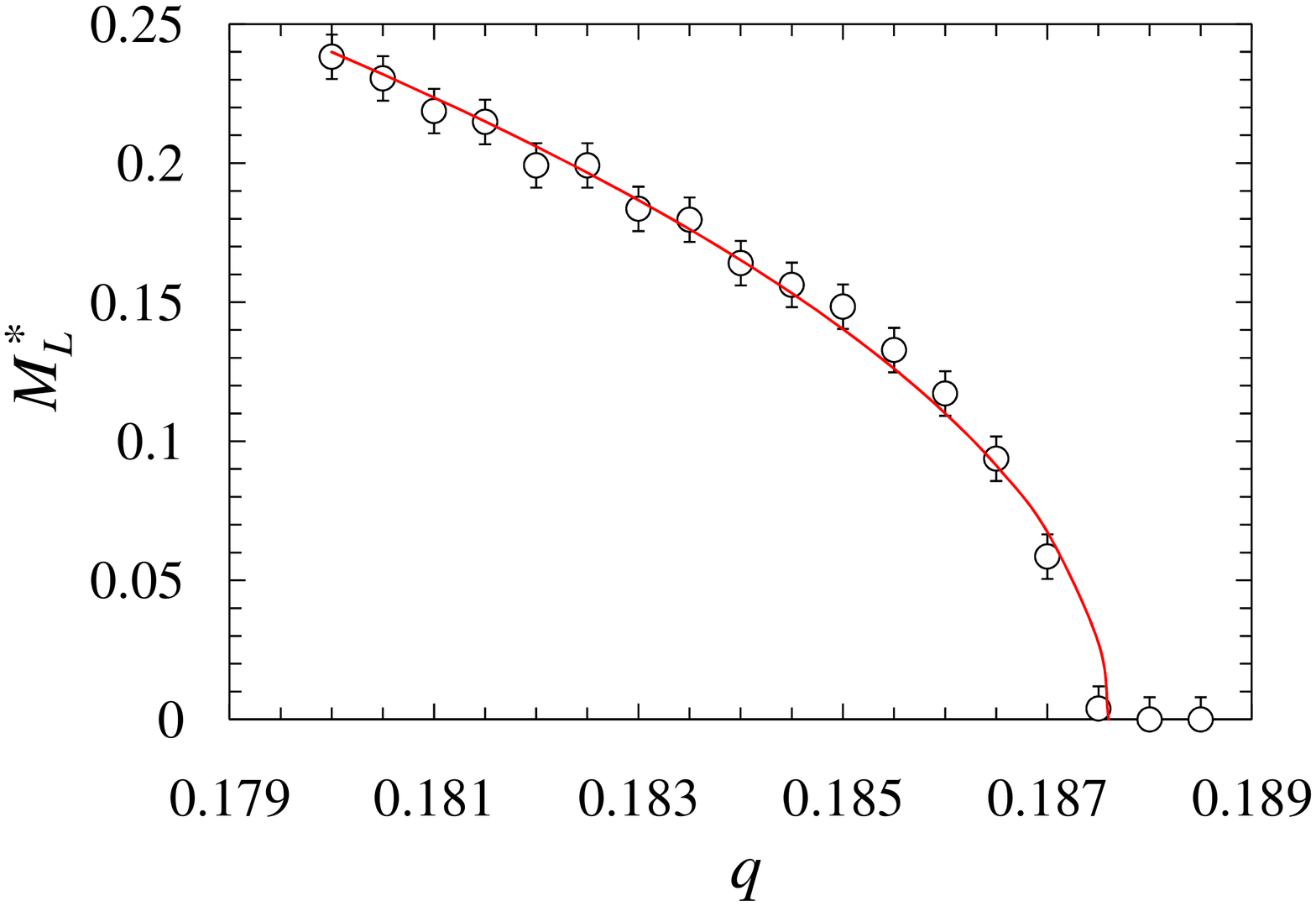}
\caption{\label{fig3}Modes of the stationary probability densities for a system with $L=512$ sites. Fits to the data (solid line) give $q_{c} = 0.1876$ and $\theta \approx \frac{1}{2}$ (see text).}
\end{figure}

We estimated $q_{c}$ and $\theta$ by fitting the modes $M_{L}^{*}$ of the probability histograms to
\begin{equation}
\label{mfit}
M_{L}^{*}(q) = A(q-q_{c})^{\theta}.
\end{equation}
We tried to use the expected values $\langle |M_{L}| \rangle = \sum_{M}|M_{L}|P_{L}(M)$ instead of the modes, but they turned out to be too spread to be useful. The precision in $M_{L}^{*}$ is bounded from below by $\delta M = 2/L$, the width of the bin of the probability histogram; in practice the uncertainties are higher because the histograms are somewhat flat and noisy at the top.

\begin{figure}
\centering
\includegraphics[viewport=68 101 518 752, scale=0.40, angle=-90, clip]{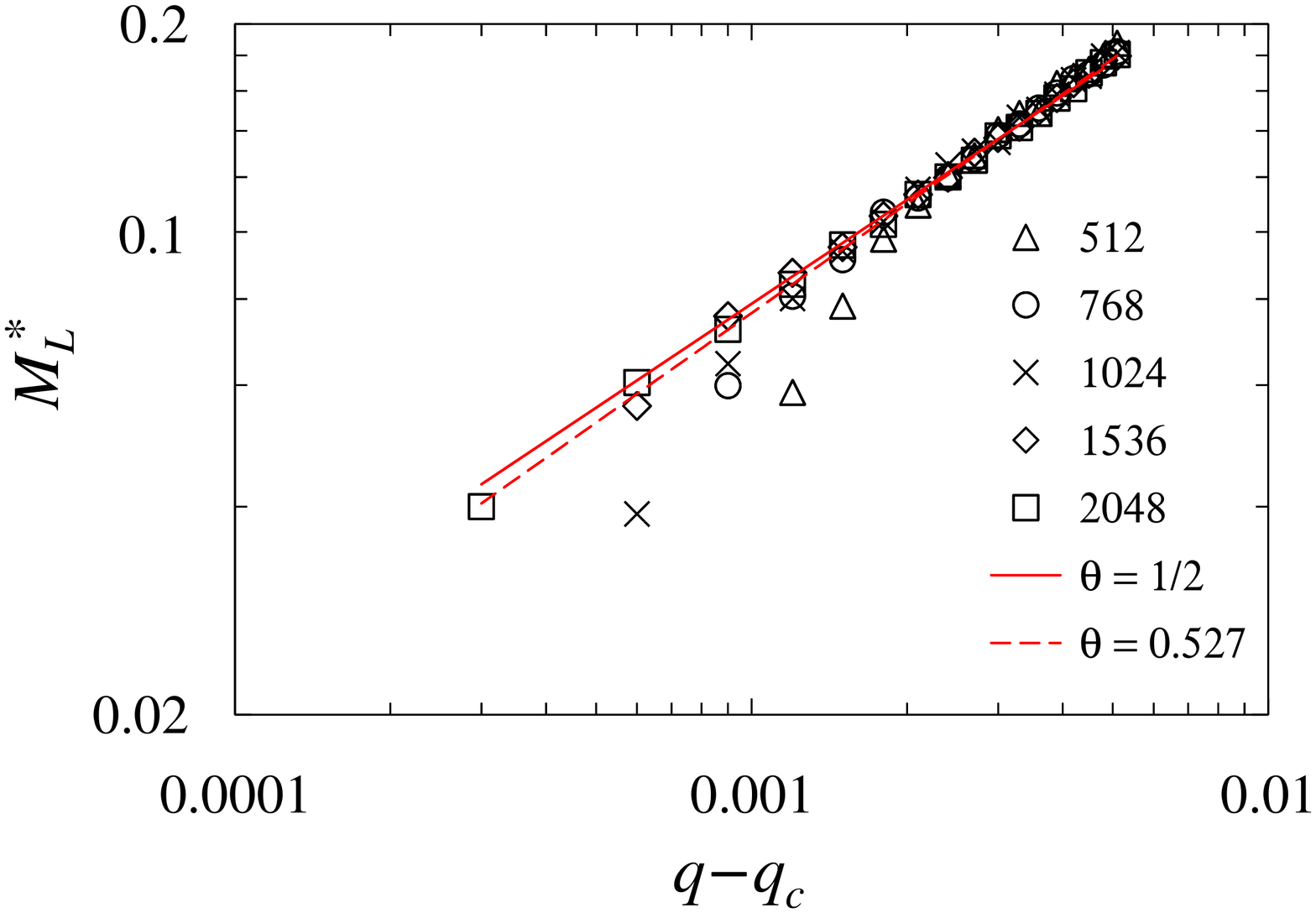}
\caption{\label{fig4}Logarithmic plot of $M_{L}^{*}$ for $512 \leq L \leq 2048$. Each point was obtained from a histogram built from $10^{7}$ samplings of $M_{L}$. The solid and dashed lines indicate the best fits to the $L=2048$ data for two different values of $\theta$ (see text).}
\end{figure}

The modes for $L=512$ are displayed in figure~\ref{fig3}. For these data, if we fix $\theta = \frac{1}{2}$ we obtain the best fit with $A=2.752$ and $q_{c}=0.1876$, with an $R^{2}=0.985$ and an adjusted $\bar{R}^{2} = 1-\frac{15}{13}(1-R^{2}) = 0.983$, and if we fix $q_{c}=0.1876$ and let $A$ and $\theta$ vary we obtain $A=2.733$ and $\theta=0.499$, with the same $R^{2}$ and $\bar{R}^{2}$ as before up to the fifth decimal place. These two fits cannot be discerned in the scale of figure~\ref{fig3} and are jointly indicated there by a solid line. Our data for larger lattices are shown in figure~\ref{fig4}. Best fits of the $L=2048$ data to (\ref{mfit}) are obtained with $A=2.489$ and $q_{c}=0.1882$ for fixed $\theta=\frac{1}{2}$, with $R^2=0.985$ and adjusted $\bar{R}^{2} = 1-\frac{16}{14}(1-R^{2}) = 0.983$, and with $A=2.903$ and $\theta=0.527$ for $q_{c}$ set at $0.1882$, now with $R^{2} = 0.999$ and $\bar{R}^{2} = 0.999$. These values of $A$, $\theta$, and $q_{c}$ are precise to the places given.

\section{\label{desorp}The model with non-null desorption from terraces}

In the model analysed so far, local smooth terraces $00$ can only evolve to dents $+-$ by adsorption of an adatom.  If we let $\Gamma^{00}_{-+} > 0$, this adsorption move gains the competition of a desorption move in which local smooth terraces $00$ lose their middle adatom and become local pits $-+$. What happens when $\Gamma^{00}_{-+} > 0$?

The model with non-null desorption rate is relevant in the study of wetting (see, e.\,g., ref.~\cite{barato} and the references there in), where an additional condition is imposed on the system, namely, that there is a wall (or floor) over which the adatoms land or depart. There cannot be desorption from the wall, and this is a very strong constraint. In our formulation this constraint does not exist.

There are many ways to take a finite $\Gamma^{00}_{-+}$. One option---the most natural, in some sense, since the rate then becomes equal to the other desorption rate involving a smooth terrace---is to take $\Gamma^{00}_{-+} = 1-q$, such that the rates $\Gamma^{-+}_{00}$ and $\Gamma^{00}_{-+}$ become `dual' to the rates $\Gamma^{+-}_{00}$ and $\Gamma^{00}_{+-}$ and the model still has a single parameter $q$. After scanning through the whole interval $0 \leq q \leq 1$ for a system of $L=1024$ sites, we did not find any sign of bimodality for $P_{L}(M)$, and we conclude that with $\Gamma^{00}_{-+} = 1-q$ the growth model is always rough. We will not provide a graph of $P_{L}(M)$ here because they are pretty featureless, relatively narrow distributions centered around $M=0$. That in this case the model is always rough is somewhat expected, because the lower the adsorption of adatoms through the $\Gamma^{-+}_{00}$ process is, the more the complementary process $\Gamma^{00}_{-+}$ digs pits in the interface, leading to an interface without any sizable smooth terrace.

Another possibility to introduce a positive $\Gamma^{00}_{-+}$ is to modify the original model (\ref{gammas}) a little more hardly by setting $\Gamma^{-+}_{00} = 1-q$ and then adding a $\Gamma^{00}_{-+} = q$. In messing this way with the original model, we must first ask whether the modified model with $\Gamma^{-+}_{00} = 1-p$ displays a phase transition at all. The answer is yes, with $\tilde{q}_{c} \sim 0.131$ and the same critical behaviour as the original model. Details about this modified model are not our concern here, though. We then proceed to modifying the original model as described above, add the desorption process with rate $\Gamma^{00}_{-+} = q$ to it, and seek a roughening transition. The result is that we did not find a roughening transition in this case either.

Finally, we can assign an independent transition rate $p \in (0,\,1]$ to $\Gamma^{00}_{-+}$. We then end up with a two-parameter model. We explored this model for small values of $p$. We found that even the smallest $\Gamma^{00}_{-+}$, as small as $p=0.001$ in a lattice of $L=1024$ sites, suffices to destroy the flat order.

As we have anticipated in the introduction (section~\ref{intro}), absence of desorption from smooth terraces provides a local mechanism that stabilises the flat phase. In the original model, the only way a flat terrace can be eroded is through desorption from its boundaries. Once we allow desorption from the middle of terraces, they loose their relative stability, particularly through the chain of reactions $+- \to 00 \to -+$, and does not survive anymore.


\section{\label{flip}Flipping times and ergodicity break}

The ergodicity of a nonequilibrium interacting particle system can be characterised by the exponential divergence of the flipping times between the different stationary states of the system as it gets larger. This divergence can be understood as the result of the superextensive growth of ``energy barriers'' between the metastable phases of the system as it gets larger, with the system getting trapped deeper inside one phase, ultimately leading to a break of ergodicity in the infinite system limit.

Here we define the flipping time $\tau_{L}(q)$ as the first time (in MCSs) it takes the initial flat surface with $M_{L}=1$ to become a rough surface configuration with $M_{L} \leq 0$. We take $M_{L}=0$ as the threshold because when the system reaches a surface configuration with $M_{L} \leq 0$ coming from a configuration with $M_{L} > 0$ it has, in an intuitive sense, ``reached the other side of the well.'' Remember that, in principle, within each sector of total charge $Q^{+}-Q^{-}$ of the stochastic process defined by rates (\ref{gammas}) all configurations are reachable and the system is ergodic.

For the purpose of exploring a possible break of ergodicity in model (\ref{gammas}), it sufices to verify that $\tau_ {L}(q)$ grows faster than algebraically with $L$. We thus look for a functional form
\begin{equation}
\label{eq:tau}
\tau_{L}(q) \sim \exp(f_{L}(q)).
\end{equation}
A nonergodic dynamics implies that $f_{L}(q)$ diverges as $L \uparrow \infty$ with finite $q < q_{c}$, while for an ergodic dynamics $f_{L}(q)$ should remain bounded in $L$. We also expect that $\tau_{L}(q) \to \infty$ as $q \downarrow 0$. This approach to determine the ergodicity of interacting particle systems has been applied successfully in several different models \cite{rakos,fuks}.

\begin{figure}
\centering
\begin{tabular}{cc}
\includegraphics[viewport=81 76 538 758, scale=0.255, angle=-90, clip]{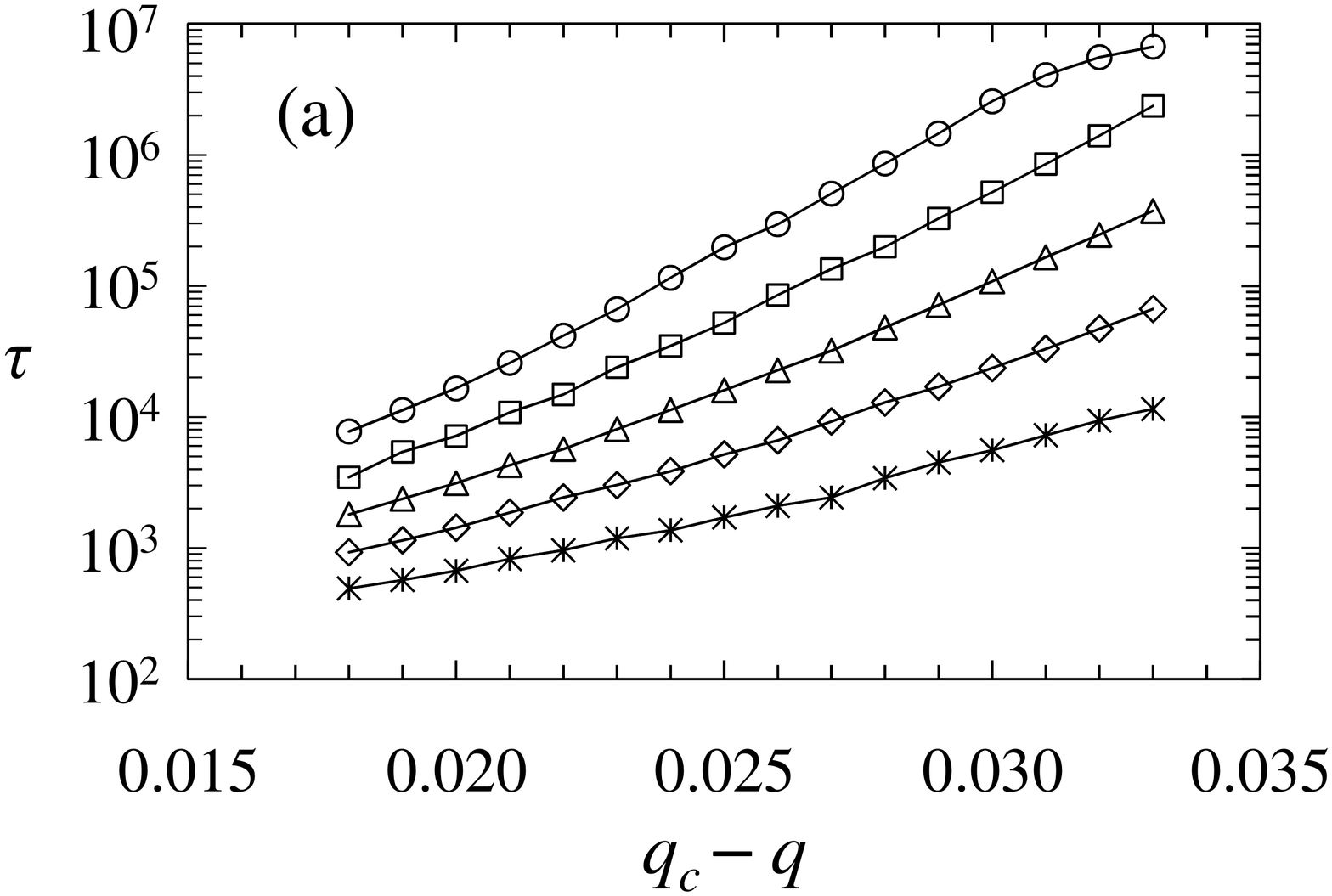} &
\includegraphics[viewport=81 76 529 748, scale=0.255, angle=-90, clip]{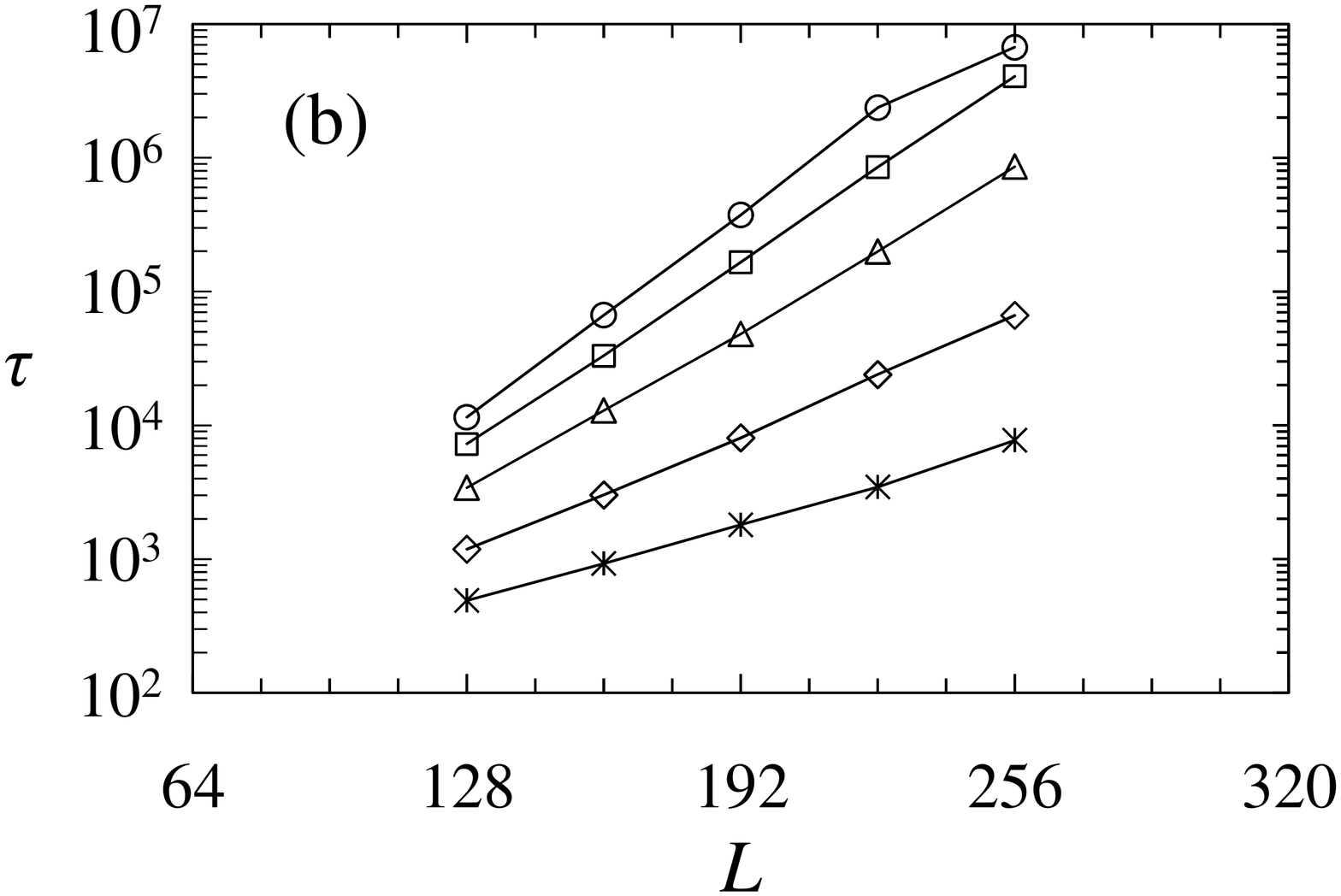}
\end{tabular}
\caption{\label{fig5}Flipping times as function of $q$ and $L$. (a)~The empirical $f_{L}(q)$ [{\it cf.\/}~(\ref{eq:tau})] grows as the system gets deeper inside the nonergodic phase $q < q_{c}$. In this figure $0.155 \leq q \leq 0.170$ and $L=128$ (lower curve), $160$, $192$, $224$, and $256$ (upper curve). (b)~In this figure, $q=0.155$ (upper curve), $0.157$, $0.160$, $0.165$, and $0.170$ (lower curve). In both cases we clearly see the exponential divergence of $\tau_{L}(q)$ as $q \downarrow 0$ and as $L \uparrow \infty$ (for $q < q_{c}$).}
\end{figure}

We initialise the flat surface with all $c_{\ell}=0$ and $M_{L}=L$, release the system and count time until $M_{L} \leq 0$. We then obtained $\tau_{L}(q)$ for each pair of $L$ and $q$ as an average over $1000$ such hitting times for some $128 \leq L \leq 256$ and $0.155 \leq q \leq 0.170$. These relatively small $L$ and $q_{c}-q$ allow us to investigate $\tau_{L}(q)$ without having to wait too much to observe the flips. Our results appear in figure \ref{fig5}. This figure clearly displays the exponential behaviour of $\tau_{L}(q)$ in the rough phase of the model, either as $L \uparrow \infty$ with fixed $q < q_{c}$ or as $q \downarrow 0$ for any $L$. This is strong indication that the model is nonergodic in the rough phase, meaning that in this phase the system does not sweep through all possible configurations compatible with the given sector of total charge $Q^{+}-Q^{-}$, getting clogged in some neighbourhood of the initial configuration. We also see from figure \ref{fig5} (and log-log plots confirm) that $f_{L}(q)$ seems to grow in a slightly superlinear fashion with $(q_{c}-q)$, i.\,e., as $f_{L}(q) \sim (q_{c}-q)^{z}$ with $z \gtrsim 1$. Our data do not allow us to estimate $z$, though.

\section{\label{summary}Summary and conclusions}

We showed that the roughening transition of the one-di\-men\-sional RSOS growth model given by rates (\ref{gammas}) occurs at $q_{c} = 0.1880 \pm 0.0004$ with an exponent $\theta = 0.51 \pm 0.02$ associated with the order parameter $M_{L}$ given by (\ref{order}). The uncertainties in these figures are conservative, because we did not perform a proper finite-size scaling analysis of the model. Our value of $\theta$ agrees with a couple of estimates published for this and related models \cite{aehm96,lopez,nnnasep}, while disagreeing with ref.~\cite{aehm98}. We believe that our data provide reasonable evidence for a mean-field exponent $\theta = \frac{1}{2}$. Incidentally, we discovered in section (\ref{desorp}) that model (\ref{gammas}) with a modified transition rate $\Gamma^{-+}_{00} = 1-q$ also displays a phase transition around $\tilde{q}_{c} \sim 0.131$ with the same critical behaviour as the original model.

An analysis of the flipping times between the metastable phases of the model in the coarse-grained ``$M_{L}$-representation'' enabled us to determine that the rough phase ($q < q_{c}$) of the model corresponds to a nonergodic phase of the stochastic process, such that the phase transition of the model can be understood as an ergodic-nonergodic transition that does not find a counterpart in one-di\-men\-sional equilibrium systems at finite temperatures.

Our results show that it is possible to investigate nonequilibrium phase transitions by sampling the probability distribution of a suitable order parameter in the nonequilibrium stationary state of the process. While histogram techniques are standard in the simulation of equilibrium systems, for nonequilibrium systems they provide a viable alternative or complement to direct time-dependent simulations.

Finally, we would like to mention that the order parameter $M_{L}$ defined in (\ref{order}) is actually one member of a family of order parameters given by
\begin{equation}
\label{mln}
M_{L}^{(n)} = \frac{1}{L} \Bigg| \sum_{\ell=1}^{L} \exp \Big( \frac{2\pi\I h_{\ell}}{n+1} \Big) \Bigg|.
\end{equation}
These order parameters can be understood as discrete Fourier transforms of the probability distribution of the heights, and are closely related with the string order parameter introduced in ref.~\cite{dennijs} to capture the disordered flat phase of two-dimensional equilibrium crystal surfaces and the spin liquid phase (a type of dilute long-range antiferromagnetic order) of one-dimensional quantum spin chains of integer spin. These parameters behave in the RSOS model like\cite{aehm98}
\begin{equation}
M_{L}^{(n)} \sim (q-q_{c})^{\theta_{n}},
\end{equation}
each with a different $\theta_{n}$. It would be interesting to estimate these exponents anew using our techniques to verify wether they assume their mean field values or not.


\section*{Acknowledgments}

The author thanks \href{http://www.mpipks-dresden.mpg.de/~saul/}{Dr.~Sa\'{u}l Ares} (MPIPKS) for bringing ref.~\cite{cuesta} to his attention and \href{http://fge.if.usp.br/~oliveira/}{Prof.~M\'{a}rio J. de Oliveira} (IF/USP) for helpful conversations. This work was partially supported by \href{http://www.cnpq.br}{CNPq}, Brazil, through grant PDS 151999/2010-4.


\appendix

\section{The `Simple KISS' pseudo-random number generator}

The `Simple KISS' pseudo-random number generator (S-KISS for short) was introduced by G.~Marsaglia in a newsgroup discussion as a simplification of his (and A.~Zaman's) `KISS' (Keep It Simple, Stupid) class of generators, which combine linear congruential, 3-shift, and multiply-with-carry generators \cite{marsaglia}. The S-KISS generator keeps only the LCG and the 3-shift parts of its parent generator.\footnote{Thus ignoring Einstein's advice to avoid making simple things simpler.} Both generators pass some stringent statistical tests of randomness collectively known as the Diehard battery of tests \cite{marsaglia}.

In 32-bit arithmetic, the S-KISS generator is implemented by the following piece of code, intended to be included in the preamble of your own main program:
{\small
\begin{verbatim}
static unsigned int A=32310901, C=1013904223;
static unsigned int X=3593214833, Y=925021637;
#define F32 (1.0/0xFFFFFFFF)
#define rnd (X=A*X+C, Y^=Y<<13, Y^=Y>>17, Y^=Y<<5, F32*(X+Y))
\end{verbatim}
}
To draw a pseudo-random number just call {\tt rnd}; e.\,g., {\tt p = rnd} attributes a pseudo-random number (purportedly) distributed uniformly in $[0,1)$ to {\tt p}. This generator has a period about as large as $2^{64} \simeq 1.8 \times 10^{19}$.

Multiplier {\tt A} is taken amongst the best for 32-bit LCG \cite{lecuyer}, while the initial values for {\tt C}, {\tt X}, and {\tt Y} are arbitrary as long as {\tt C} is odd and {\tt Y} is non-null. Our initial {\tt X} and {\tt Y} have an equal number of bits {\tt 0} and {\tt 1}, not in any obvious pattern, just in case. You can change your favorite 32-bit LCG for the one given above. The shifts in the 3-shift part of the PRNG (the numbers {\tt 13}, {\tt 17}, and {\tt 5}) as well as their order are not arbitrary; for other possible triples consult ref.~\cite{marsaglia}.

Neither the KISS nor the S-KISS generators were designed for cryptographic applications. This relative limitation notwithstanding, we believe that their simplicity, quality, high throughput, and large enough periods make them well suited for scientific applications like our Monte Carlo simulations. In fact, in this regard it would be interesting to examine these pseudo-random number generators using physics-based tests to assess their reliability in statistical physics simulations \cite{tapio,codd}.


\end{document}